\documentstyle[12pt,epsf]{article}
\catcode`\@=11 \@addtoreset{equation}{section}  
\renewcommand{\theequation}                     
         {\arabic{section}.\arabic{equation}}   
\title{Inertia as the ``Threshold of Elasticity'' 
of Quantum States}
\author{P. Leifer}
\date{School of Physics and Astronomy, Raymond and Beverly Sackler Faculty of Exact Sciences \\ Tel-Aviv University, Tel-Aviv 69978, Israel}
\begin{document}
\maketitle
\begin{abstract}
The principle of general relativity (GR) of Einstein is based on
the possiblity ``to create'' or ``to compensate'' 
locally the gravitation field by a choice of the 
accelerated frame
in spacetime. 
In quantum field theory do not exist pointwise
particles--``material points'' (what we need for the locality), but we have rather extended wave packets 
(quantum particles)
with ``polarizations'' of different kinds that 
correspond, in principle, to indefinite number of 
internal degrees of freedom.
In this case both the spacetime manifold is very poor 
and
the factorization of all motions in accordance with 
the classification--``uniform/accelerated''
is very rough, for the desirable possibility
``to create'' or ``to compensate'' any physical
field. Hence, we should deal 
with a quantum state space. Therefore, it is natural 
to think that 
creation and compensation (annihilation) of the 
quantum field may be related to the
choice of the functional frame in the state space 
and that 
must be some new principle of classification of quantum motions. This hypothesis has been called ``superrelativity''. This leads to a new gauge theory 
of the geometric type in the projective Hilbert space.
 
Here I try to connect the inertia property of 
``elementary'' quantum particles with the  deformation 
of quantum modes which prevent these particles 
from the flying apart and to correct some formulas
of my previous articles.
\end{abstract}
\vskip .2cm
\section{Introduction. The Quantum Origin of Inertia}
Einstein stated in his attempt to constitute the principle of the {\it general covariance} that the reason
of the difference in the behavior of the two liquid
bodies $S_1$ and $S_2$ lies {\it outside of the system,
namely--in the system of distant stars}.
I think, however, that here is some confusion since
the system of the distant stars in this discussion
takes place of the {\it absolut reference frame as a true
discriminator} of the character of motions in spite
of Einstein statement that under a correct formulation 
of  physical laws, any frame must be available 
\cite{Einstein}. 
It is not enough, however, to conclude that there is 
some inconsequence in the foundations of {\it classical} general relativity. Furthermore, I think that even
the {\it notion of ``acceleration'' should be avoided
in a consistent quantum gravity} because, for instance,
the physical state of a free falling down droplet of 
mercury in gravitational field is indistinguishable 
from the physical state of the similar droplet which is 
far from masses. If and only if there exists some acceleration force which evokes changing of the physical state of the droplet (temperature, surface tension etc.), one can think
about ``accelerated motion'', although in particular case
the droplet may even rest relative, say, surface of a 
table.
This example shows  that instead of indefinite
notion of ``acceleration'', an immanent characteristic 
of the spacetime motion should be used--internal
(as a matter of fact--quantum) states of system. 
\footnote{As far as I know only in the two letters  from Schr\"odinger to Einstein (18.11.50) and from Einstein
to Schr\"odinger (22.12.50) the problem of acceleration in quantum
theory has been discussed. Einstein wrote in his letter
that ``full description can not be built on the notion of 
acceleration''.} Hence,  we can point out the physical
reason of the different behavior of the bodies $S_1$ and 
$S_2$, if we will take into account a {\it physical
state of these bodies}. But, if the distant stars
can not play the role even of a ``true'' reference frame
for the measurement of accelaration, how one can think
that they are a source of the masses of ``elementary''
particles, droplets of mercury etc.?
We have to find some different source of the inertia
proprties of matter.

How we can do it? Now it is well known that the 
existence and stability of extended macroscopic objects 
are supported by Goldstone's modes \cite{TFD}. 
{\bf I think we can connect the inertia property 
rather with the quantum conditions of excitations and reconstruction of these modes} \cite{Le7} than with 
the Mach principle \cite{Einstein,Einstein2,G}. Then, 
in some sense, one can say that the source of inertia 
lies, in fact, ``outside'' the spatial boundary of body,
namely--in the Hilbert state space. This means that the deformation of the self-consistent boundary leads to the generation of the collective mode as a dynamical quantum reaction on 
external force. {\bf Therefore an  acceleration is 
only external (macroscopic) ``exibit'' of internal 
quantum reaction \cite{Le5,Le6,Le7}. That is one 
may treat the inertia as an ``elasticity'' 
of the self-consistent boundary at the quantum level}.
The deformations of this boundary leads to deformations
of the quantum states of electron's shells
and these excitations one can interpret as evidence
of the inertial properties of a physical system.

Here we try to connect  the inertia property of an ``elementary particle''  and the deformation of the 
internal quantum collective modes which prevent this 
system from flying apart under some kinds of repulsive forces. Dirac applied a surface tension force of some 
non-Maxwellian type as a stabilizer  in the extensible 
model of the electron \cite{Dirac1}.
K.R.W. Jones pursuing to reach in his beautiful works \cite{Jones1,Jones2} the objective interpretation
of the wave function, have used the Newtonian
gravitational self-energy in order to find bound and
stable solitary wave solution of the nonlinear 
{\it gravitational Schr\"odinger equation} of the
scalar Bose-Einstein condensate. That is  
this nonlinearity should play the role of the physical mechanism of the suppressing  of the dispersion which 
leads to spread out wave packets. I think that for
the relativistic realization of these ideas, a
geometrical approach is unavoidable and that geometrical spirit of Einstein program should be conserved.
{\bf Instead of nonrelativistic scheme of Jones 
I propose to use the superrelativity principle which 
returne us to the geometrical origin (in projective 
Hilbert space) of the unified interactions in 
accordance with Einstein's program. Then the curvature 
$R^s_{kj^*i}$ of the CP(N) is the quantum source of the inertia of the nonlocal 
elementary particles}. Thereby the self-interacting 
fields
of Barut's type \cite{Barut} has a geometric origin.
They arise as a state-dependent gauge fields ``of the
third kind'' of Doebner-Goldin \cite{Goldin}.

\section{The Physical Quantum Projective Space}
I had already discussed the dynamical sense of 
the projective Hilbert space and the Fubini-Study 
metric \cite{Le1,Le2,Le3,Le4,Le5,Le6,Le7}. Of course,
any appropriate choice of the local coordinates is acceptable, but $\Pi$ coordinates is as natural as, 
for example, the Cartesian coordinates in Euclidean 
space. The using of CP(N) geometry is closely connected 
with both the new point of view on the 
role of spacetime structure and physical field concept.
\subsection{Spacetime and Field Concept}   
It has been mentioned early \cite{Le1,Le3,Le5,Le6} that 
we should sacrifice the priority of spacetime manifold 
and to build quantum field theory over state space.
From the technical point of view this necessitate a 
new construction of {\bf quantum physical fields over
state space. Only after that the spacetime 
propagation of their ``envelopes'' should be 
described}.  

When we build our theories we usualy begin with
some fields on spacetime manifold. Thus, we think
that it is the most realistic approach to the 
achievement of an agreement between theory and 
experiment. But we should remember 
that the measurement of the spacetime position of the ``event'' is {\bf only the ``label'' of the fact 
of some quantum transition in a quantum state space}
\cite{Le1,Le2,Le3}. Therefore
we can assume that from the physical point of view 
just the quantum dynamics in the state space is 
interesting for us. Hence, it is more natural to 
deal with functions on quantum
state space than with the ordinary spacetime functions (classical or ordinary quantum fields).

That is, instead of a material point and functions of
its coordinates in spacetime, one should use functions of the coordinates of a quantum state. These tensor fields 
take place of dynamical variables which have reasonable
analitic properties. Otherwise
we obtain singular results because  a pointwise (in spacetime) dynamical variables contain singular functions. 
In fact, the all known variants of the regularization procedure are the processes of a ``delocalization'' of 
the pointwise dynamical variables. {\bf Thus, in our 
case we search functions of ``quantum transition'' in the space of the ``internal'' degrees of freedom, not 
functions of coordinates of the ``event'' in spacetime 
which
(coordinates) are, after all, only obtrusive illusion}. Dynamical (habitual) spacetime 
arises only as a manifold of the ``centrum of mass'' of quantum ``droplets''--the soliton-like solutions of some 
nonlinear wave equations. 

I would like to emphasize the fact that what was 
historically called  ``coupling constants'' are, 
as a matter of fact, {\it spacetime functions} under renormalization procedure. 
For example, the Newton's constant $G_N$ should be 
subjected
a renormalization in quantum area because, if the 
mass of some particle $m$ corresponds to  
the oscillation process
with the frequency $\omega $ by the equation 
$mc^2=\hbar \omega $, it should have (in accordance 
with the Einstein's formula for a frequency) a larger
value--``blue shift of de Broglie wave'' in the 
vicinity of a massive body M
\begin{equation}
m_0=\frac{\hbar \omega_0}{c^2}=\frac{\hbar \omega}{c^2}(1+\frac{G_N M}{c^2 d}),
\end{equation}
where $\frac{G_N M}{d}$ is a gravitational potential of 
a massive body. This result was interpreted by Einstein
as an affirmation of Mach principle \cite{Einstein2}.
Thus, the Mach mechanism of mass generation must have
fantastic, almost infinite resolution
$r_0/L_G=10^{-13}cm/10^{20}cm=10^{-33}$ for
the spatial separate generation of quite different 
masses of elementary particles at nuclear distance.
It is defined by the relation of the nuclear size to 
the remote galaxy distance. Besides that, on the 
quantum level it leads to
the infinite self-interacting gravitation field 
which we can not see in experiments.
Similar arguments are applicable to the electroweak and 
to the set of strong coupling ``constants''.
That is {\it any field construction which based on 
spacetime dependent functions (fields) and pointwise
charges as sources of these fields leads to difficulties
and requires renormalization of these charges.}
Therefore one have to use a new set of primordial
elements in the quantum area \cite{Le1,Le5,Le6}.
The program of the overcoming of these obstacles
must be based on the isometry between the 
coset structure of quantum interactions and the 
projective quantum state space CP(N).

\subsection{Coset Structure of Universal 
Quantum Interactions}
As was mentioned early \cite{Le5,Le7} the physical 
deformations of the pure quantum state have the geometric 
structure of a coset, i.e. the structure of 
the CP(N): $G/H=SU(N+1)/S[U(1)_{el} \times U(N)]=CP(N)$. This paves the way to the invariant
study of the spontaneously broken unitary symmetry
\cite{Le5}.
This statement has a general character and does not 
depend on particular properties of the pure quantum
state. {\bf The reason  for the change of motion of 
material point is an existence of a force. The reason 
for the 
change of a pure quantum state is an interaction which 
is evoked by unitary transformations from the coset
G/H. The reaction of a material point is an acceleration. The reaction of a pure quantum state is its deformation
whose geometry is in fact the geometry of the 
coset transformations $SU(N+1)/S[U(1)\times U(N)]$.
That is a universal quantum interactions at the 
fundamental level are rooting in the Hilbert projective state space CP(N)}.

These simple speculations can help presumably to solve 
the problem of hierarchy of the fundamental interaction
and masses. The fine structure constant $\alpha$ in the Lorentz-radial Klein-Gordon equation (in the Lommel form) \cite{Le4,Le6} takes place of the {\it dimensionless
mass (or energy) of a fundamental scalar field}. 
The geodesic excitations of this scalar field
which are induced by transformations from the 
coset should be somehow related to the spectrum of masses
of nonlinear particle-like solution of nonlinear wave
equation which arises under the geodesic variation
of the Fourier components of the solution of 
the Lommel equation \cite{Le4,Le5,Le6}.
It was be shown \cite{Le6,Le7} that excitations
which was generated by the geodesic flow in CP(N)
correspond to quite concrete elements of the invariant
subspace in the algebra AlgSU(N+1)
\begin{equation}
\hat{B}= \left(
\matrix{
0&f^{1*}&f^{2*}&.&.&.&f^{N*} \cr
f^1&0&0&.&.&.&0 \cr
f^2&0&0&.&.&.&0 \cr
.&.&.&. &.&.&. \cr
.&.&.&. &.&.&. \cr
.&.&.&. &.&.&. \cr
f^{N}&0&0&.&.&.&0
}
\right ).
\label{generator} 
\end{equation}
as functions of the Fourier components of the initial 
state (solution of the Lommel equation)
$|f^i|=g|\Phi^i|(R^2-|\Phi^0|^2)^{-1/2}$, where 
$\arg f^i=\arg \Phi^i$ (up to the general phase),
and angle from the equation $cos\Theta=|\Phi^0|/R$.
This is a natural
type of the ``second quantization'' which has,
however, dynamical character \cite{Le7}. That is,
instead of the Dirac's {\it identification} of Fourier
components with operators which obey Fermi or Bose
commutation rules \cite{Dirac2} we have a {\it invariant correspondence} between them. This ``second 
quantization'' can not be related to some ensemble 
of the identical quantum particles
but rather to internal oscillations near the bottom
of the valley of the intrinsic quantum potential
of the natural connection in CP(N) 
\begin{equation}
\Gamma^i_{kl} = -2 \frac{\delta^i_k \Pi^{l*} + \delta^i_l \Pi^{k*}}{R^2 + \sum_{s=1}^{N} |\Pi^s|^2},
\label{connection} 
\end{equation}
\cite{Le5,Le6,Le7}.
{\it The physical unity of theory demands only single true fundamental interaction constant $\alpha=\frac{e^2}{\hbar c}$, that, from the geometric point of view, is the sectional curvature of the CP(N)} $\kappa=\alpha=R^{-2}$.
Then all physical coupling constants should arise
as charectiristics of the deformation of quantum state
in different directions of the quantum state space
CP(N) under action of the coset transformations
in the form $\frac{f^i f^{*k}}{g^2}(\cos \Theta -1)$.
This requires not only topological equivalence of the 
coset manifold $SU(N+1)/S[U(1)\times U(N)]$ and CP(N). 
There is isometry between these manifolds \cite{KN}.
In order to identify the metrics of the coset manifold 
and CP(N) it is enough to identify the tangent space 
$T_0SU(N+1)/S[U(1)\times U(N)]$ which containes matrices
F,G like (\ref{generator}) and the tangent space $T_0CP(N)=C^N$ which containes vectors $(f^1,f^2,...,f^N),(g^1,g^2,...,g^N)$. 
Then one should define the ``Lagrangian of the geodesic
velocities of deformation'' as a scalar product
\begin{equation}
L_{fg}=(F,G)=\frac{1}{2}Tr FG^+=Re(f,g)_{C^N}=Re \sum_{i=1}^N f^i g^{i*}.
\end{equation}
In order to establish the isometry globaly over whole
group we have to introduse the Killing metric
by using left-invariant vector fields $A_F(u)=uF,
B_G(u)=uG, u \in SU(N+1), F,G \in AlgSU(N+1)$.
The Killing scalar product is
\begin{equation}
<A,B>=Re Tr A B^+|_0
\end{equation}
for $A,B \in SU(N+1)$ is $Ad[U(N+1)]$-invariant. 
Then $r^2=||g||^2=<g,g>=Re Tr E=N+1$.
This Killing metric should be in the ``harmony'' with the 
Fubini-Study metric in CP(N). That is
SU(N+1) and the coset are embedded into the sphere 
with the radius $r=\sqrt{N+1}=\sqrt{2S+1}$, where 
multilevel $N+1$ system (the multiplet of excitations,
for example) may be expressed in terms of spin $S$. 
The radius of the sectional
curvature of CP(N) is, in accordance with our 
assumption, $R=\alpha^{-1/2}$. 
Thus, the radius $R$ has a physical meaning. Therefore
we should normalize correspondence between R and r. 
Thereby, we have here some kind of the {\bf superselection 
rule} for the ``mass--spin'' relationship.
The question, of course, is, whether physically correct
the identification of S with the spin of particles. 

Thereby physical effects of a quantum 
dynamics in CP(N) are curvature-dependent. The 
projective symmetry is broken. The unitary symmetry 
is local and hidden. In this case,
what variables $\Psi$ say about? Naturaly think 
that they describe not amplitudes of probability of some 
ensemble,
but they should be ``classical'' Fourier amplitudes 
of an internal quantum dynamics of a single particle. 
This idea corresponds to the Barut's approach to the nonperturbative ``self-field'' version of the
quantum electrodynamics \cite{Barut}.  These related
to the Goldston and Higgs excitations of the geometric type
which arise under deformation of quantum states in the
self-interacting potential of the local unitary rotations
of the functional frame in the Hilbert space
\cite{Le7}. It is some analog of the ``string excitations''
if somebody like to think in the framework of this model.
Then one can say that the geodesic of the Fubini-Study metric plays the role of a closed string \cite{Le6}. 
{\it But, of course, this mechanism has a quite different physical meaning}. I intend to develop here the 
differintial-geometric aspects of a gauge theory over 
CP(N) with the metric tensor  in the local coordinates 
$\Pi$  
\begin{eqnarray}
G_{ik*}=R^2 \left[\frac{(\sum_{s=1}^N |\Pi^s|^2+R^2)\delta_{ik}-
\Pi^{*i}\Pi^k}{(\sum_{s=1}^N |\Pi^s|^2+R^2)^2}\right].
\end{eqnarray}

\section{The Geometric Origin of Elasticity of 
Quantum States}
In our approach, the projective 
Hilbert space CP(N) of the generalized coherent states is 
treated as the {\it configuration space of wave packets}. That is, 
we regard the Fourier components of the scalar (real or complex) 
classical field as a ``multiplet'' relative to $SU(N+1)$ 
symmetry broken down to the isotropy group 
$U(1) \times U(N)$ of the ``vacuum'' vector. 

{\it The essentially a new element of our approach is the 
action of geodesic flow in the configuration space on the relative Fourier components $CP(N)$ 
\cite{Le6}. The principle of least action arises in 
$CP(N)$ as a principle of minimal distance. The integral curves of the geodesic flow (geodesics) are stable and closed (periodic)}. 

{\bf The key idea proposed here associated with a model for a quantum particle as an extended field dynamical system. Internal ``hidden'' coordinates of a ``pre-system'' is local coordinates in the projective Hilbert space CP(N). Self-evolution of this ``pre-system'' appears as a motion along closed path in CP(N). This path is a geodesic of the Fubini-Study metric. The positive curvature of the Hilbert projective space is the reason for the stability of geodesics in CP(N). We interpret this internal local
coordinates as Fourier componets of some wave packet
in the reference Minkowski spacetime. Therefore the ``pre
-system'' is ``wrapped'' in the shell of a surrounding field in ordinary spacetime and one has, hence, a non-local
extended dynamical field configuration--``droplet''}.

The flow is then given by the unitary matrix
$\hat{T}(\tau,g)=\exp(i\tau\hat{B})=$
\begin{equation}
\left(
\matrix{
\cos\Theta&\frac{-f^{1*}}{g} \sin\Theta&.&.&.&\frac{-f^{N*}}{g}\sin\Theta \cr
\frac{f^1}{g}\sin\Theta&1+[\frac{|f^1|}{g}]^2 (\cos\Theta -1)&.&.&.& \frac{f^1 f^{N*}}{g^2}(\cos\Theta-1) \cr
.&.&.&.&.&.\cr
.&.&.&.&.&.\cr
.&.&.&.&.&.\cr
\frac{f^{N}}{g}\sin\Theta&\frac{f^{1*} f^{N}}{g^2}
(\cos\Theta-1)&.&.&.&1+[\frac{|f^{N}|}{g}]^2 (\cos\Theta -1)
}
\right),
\label{flow}
\end{equation}
where $g=\sqrt{\sum_{k=1}^{N}|f^k|^2},\Theta=g\tau$ 
\cite{Le1,Le5}.
The form of the periodic geodesic ``deformation'' of the  Fourier components of the initial solution 
\begin{eqnarray}
\Phi^a=\frac{a!}{\Gamma (a+1/2)}\int_0^{\infty}
y^{-1/2}L_a^{-1/2}e^{-y}J_1(\alpha \sqrt{y})dy.
\end{eqnarray}
of the Lommel equation
\begin{equation}
\frac{d^2 \Phi^*}{d \rho^2} + (3/\rho) \frac{ d \Phi^*}{d \rho} + \alpha^2 \Phi^* = 0, 
\label{lommel}
\end{equation}
is represented by the formula 
\begin{equation}
|\Psi (\tau,g,y)> = \sum^{N}_{a,b=0} \Phi^a 
[\hat{T}^{-1}(\tau,g)]^b_a  |b,y> ,
\label{droplet}
\end{equation}
where in the particular case $\tau=0$ one has
\begin{equation}
|\Psi (0,g,y)> = \sum^{N}_{a,b=0} \Phi^a 
[\hat{T}^{-1}(0,g)]^b_a  |b,y> 
=\sum^{N}_{a,b=0} \Phi^a \delta ^b_a  |b,y> =
|\Phi>,
\label{phi}
\end{equation}
and for $\tau 'g'=\Theta '$ when $\cos\Theta '=|\Phi^0|/R$ 
\begin{eqnarray}
|\Psi (\tau ',g',y)> = \sum^{\infty}_{m,n=0} \Phi^m 
[\hat{T}^{-1}(\tau ',g')]^b_a  |b,y> \nonumber \cr 
=\sum^{N}_{a,b=0} R \exp [i\omega (\Phi)] \delta ^b_0 |b,y> = R \exp [i\omega (\Phi)] |0,y> =|\Psi_0>.
\label{psi}
\end{eqnarray}
Thus direction of the transformation of the 
vacuum vector into the solution of the Lommel equation is 
determined by the matrix 
$\hat{P}=\hat{G}^{-1}(\Phi)\hat{B}(\Phi)\hat{G}(\Phi)$,
that is all algebra $AlgSU(N+1)$ subjected to the 
proper (in the Hilbert space) similarity transformation 
for the adaptation of the algebra structure to any state vector.

Briefly speaking, one can define a proper state-dependent 
images of the Cartan's decomposition of
the elements of Lie algebra $AlgSU(N+1)$
$h_\Phi=\hat{G}^{-1}(\Phi)\hat{h}\hat{G}(\Phi)$, and
$b_\Phi=\hat{G}^{-1}(\Phi)\hat{b}\hat{G}(\Phi)$.
It is easy to proof that the proper commutation 
rules are commonly known:
\begin{equation}
[h_\Phi,h_\Phi]\subset h_\Phi,\quad [b_\Phi,b_\Phi]\subset h_\Phi,\quad [h_\Phi,b_\Phi]\subset b_\Phi.
\label{prophb}
\end{equation}
But such approach is neither convenient nor logical
for our purposes because on every step of the ansatz
of sqeezing one should to solve the ``elimination equations'', and this approach appeals an artificial   
``vacuum'' state 
\begin{equation}
|\Psi_0>= \left(
\matrix{
e^{i\omega(\Psi)} \sqrt{\sum_{a=0}^N |\Psi^a|^2} \cr
0 \cr
. \cr
. \cr
. \cr
0 
}
\right )= R e^{i\omega(\Psi)}\left(
\matrix{
1 \cr
0 \cr
. \cr
. \cr
. \cr
0 
}
\right ),
\label{vac} 
\end{equation}
which is not in general
the real vacuum state of a concrete physical problem.
The most consistent approach is {\it the using of the 
local coordinates}
\begin{equation}
\Pi^1=R\frac{\Psi^1}{\Psi^0},\quad \Pi^2=R\frac{\Psi^2}{\Psi^0},...,
\Pi^i=R\frac{\Psi^i}{\Psi^0},...,
\Pi^N=R\frac{\Psi^N}{\Psi^0},...
\label{Pi}
\end{equation}
\cite{Le1,Le2,Le3}.
That is, we now refer to 
the term "local" as a fact of a dependence on the 
local coordinates in the CP(N).
One should find the relationship between the linear
representations of SU(N+1) group by an  
``polarization operator'' 
$\hat{P}=\mu H^{\sigma} \hat{\lambda}_{\sigma} \in AlgSU(N+1)$ 
which depends on external ``multipole magnetic'' 
or ``gluon'' field $H^{\sigma}$,
$1 \leq \sigma \leq N^2+2N$ and does not depend 
on the state of the quantum 
system, and the  nonlinear representation 
(realization) of the group symmetry in which the infinitesimal operators of the transformations depend 
on the state.  In the linear representation of the 
action of $SU(N+1)$ we have
\begin{equation}
|\Psi(\tau)>=\exp(-\frac{i}{\hbar}\tau \hat{P}) |\Psi>.
\end{equation}
For a full description of a group dynamics by pure 
quantum states, we shall  
use coherent states in CP(N). If $\hat{P}_\sigma$ is
one of the $N^2+2N$ directions in the 
group manifold one has
\begin{equation}
D_\sigma(\hat{P})=\Phi^i_\sigma (\Pi,P)\frac{\delta}{\delta \Pi^i}
+\Phi^{i*}_\sigma (\Pi,P)\frac{\delta}{\delta \Pi^{i*}},
\end{equation}
where
\begin{equation}
\Phi_{\sigma}^i(\Pi;P) = R\lim_{\epsilon \to 0} \epsilon^{-1}
\biggl\{\frac{[\exp(i\epsilon P_{\sigma})]_m^i \Psi^m}{[\exp(i \epsilon P_{\sigma})]_m^k
\Psi^m }-\frac{\Psi^i}{\Psi^k} \biggr\}=
\lim_{\epsilon \to 0} \epsilon^{-1} \{ \Pi^i(\epsilon P_{\sigma}) -\Pi^i \}
\end{equation}
are the local (in CP(N)) state-dependent components  of the SU(N+1) group generators, which are studied in \cite{Le1,Le2,Le3}.

In order to establish relationships  between ``internal'' parameters in CP(N) and a propagation of the 
scalar field near the light cone in the 
``reference spacetime'', we should ``lift'' a geodesic deformation of the initial Fourier components into the 
fiber bundle. Namely, if we assume that {\it in accordance with the ``superequivalence principle'' \cite{Le5} an infinitesimal geodesic ``shift'' of field dynamical variables could be compensated by an infinitesimal transformations of the 
basis in Hilbert space, then one can get some effective self-interaction potential as an addition  
term in original Klein-Gordon equation in the Lommel 
form} (\ref{lommel}).  
We will label hereafter vectors of the Hilbert space by Dirac's notations
$|...>$ and tangent vectors to CP(N)  
(field dynamical variables) by 
arrows over letters, $\vec \xi \in T_{\Pi}CP(N)$, 
for example. Then one has a definition
of the rate of a state vector  changing
\begin{equation}
|v(y)>=-(i/\hbar)\hat{P}|\Psi (y)>.
\label{rate}
\end{equation}
The ``descent'' of the vector field $|v(y)>$ onto the 
base manifold 
CP(N) is a mapping by the two formulas:
$f:\cal H \rm \to CP(N)$, i.e. (\ref{Pi})
\begin{eqnarray}
f:(\Psi^0,...,\Psi^i,...,\Psi^N) \to (R\frac{\Psi^1}
{\Psi^0},..., R\frac{\Psi^N}{\Psi^0},...)=
(\Pi^1,...,\Pi^N),
\label{map1}
\end{eqnarray}
and
\begin{eqnarray}
\vec \xi=f_{*(\Psi^0,..., \Psi^N)} |v(y)> 
=\frac{d}{d \tau}(R \frac{\Psi^1}{\Psi^0},..., R \frac{\Psi^N}{\Psi^0})\Bigl|_0 \cr  
=\frac{d}{d \tau}(\Pi^1,...,\Pi^N)\Bigl|_0
=-(i/\hbar)[R P^1_0-P^0_0 \Pi^1+(P^1_k-(1/R)P^0_k \Pi^1)\Pi^k,...,\cr
R P^N_0-P^0_0 \Pi^N +(P^N_k-(1/R) P^0_k \Pi^N)\Pi^k].  
\label{map2}
\end{eqnarray}
That is the operator $\hat{P}$ determines a field 
dynamical variable $\vec \xi$ (\ref{map2}). On the 
geodesic which spans both the vacuum vector (\ref{vac}) 
and the solution of the Lommel equation there is a ``natural'' vector field $\vec \xi(\Pi(\tau))$ which, of course,
is, in general, non-parallel along the geodesic.
One can look on the integral curve of this vector field
as on some ``excited string'' in CP(N) under a
perturbation of the ``geodesic string''.
We propose  the following guide of realization of 
``superequivalence'' principle:

1. To use the covariant derivative of the vector field 
$\vec \xi(\Pi(\tau))$ relative the Fubini-Study metric
in order to keep the tangent vector field. It is well 
known that at a point  
$\Pi+\Delta \Pi$ in $CP(N)$ the ``shifted'' field  
$\vec \xi+ \delta \vec \xi$
$=\vec \xi+\frac{\delta \vec \xi}{\delta \tau} \delta \tau$
contains the derivative $\frac{\delta \vec \xi}{\delta \tau}$, which is not, in the
general case, a tangent vector to CP(N), but the 
{\it covariant derivative}
$\frac{\Delta \xi^i}{\delta \tau}$
$=\frac{\delta \xi^i}{\delta \tau}+\Gamma^i_{km}
\xi^k \frac{\delta \Pi^m}{\delta \tau}$
{\it is} a tangent vector to CP(N).

2. To make a small shift along the covariant derivative
$\frac{\Delta \xi^i}{\delta \tau}\delta \tau$.

3. It may be proved that for the shape of the ellipsoid of polarization at the point $\Pi+\delta \Pi$
along the direction of the covariant derivative
$\frac{\Delta \xi^i}{\delta \tau}\delta \tau$
on the geodesic from $\{0\}$ to $\Pi$ there is
a point $\Pi + \Delta \Pi$ where ellipsoid of 
polarization has 
the same shape. That is instead of the shift along
integral curve of the vector field
$\vec \xi(\Pi(\tau))$ one can use a shift along
the geodesic which was mentioned above.
Thereby one can avoid dufficulties with zeroth
modes which arise under the general variation involving
tarnsformation from the isotropy group. 

4. Now we should ``lift'' the new tangent
vector $\xi^i + \Delta \xi^i$ into the original Hilbert space $\cal H \rm$,
that is, one needs to realize two inverse mappings: 
$f^{-1}:CP(N) \to \cal H \rm $ at point $\Pi^i+
\Delta \Pi^i$ by the formula 
\begin{equation}
\Psi^{'0}=\frac{R^2}{\sqrt{\sum_{s=1}^N |\Pi^s+\Delta \Pi^s|^2+R^2}},...,
\quad 
\Psi^{'i}=(\Pi^i+\Delta \Pi^i) \frac{R}{\sqrt{\sum_{s=1}^N |\Pi^s+\Delta \Pi^s|^2+R^2}}.
\label{PsiPi'},
\end{equation}
or in the first approxipation
\begin{eqnarray}
f^{-1}:(\Pi^1+\Delta \Pi^1,...,\Pi^N + \Delta \Pi^N) 
\to [\Psi^0+\frac{\partial \Psi^0}{\partial \Pi^i} 
\Delta \Pi^i ,...,
\Psi^N+\frac{\partial \Psi^N}{\partial \Pi^i}\Delta \Pi^i].
\label{map-1}
\end{eqnarray}
and then
\begin{eqnarray}
f^{-1}_{* \Pi+\delta \Pi}(\vec \xi + \Delta \vec \xi)
=[v^0+\Delta v^0,v^1+\Delta v^1,...,v^N+\Delta v^N] \cr
=[\frac{\partial \Psi^0}{\partial \Pi^i}(\xi^i+
\Delta \xi^i),
 \frac{\partial \Psi^1}{\partial \Pi^i}(\xi^i+\Delta \xi^i) ,...,
\frac{\partial \Psi^N}{\partial \Pi^i}(\xi^i+\Delta \xi^i)].
\label{map-2}
\end{eqnarray}
It is may be shown that under the parallel transport
of the  $\vec \xi$ along a smooth curve, one has
\begin{equation} 
\Delta \xi^i=\xi^i(\tau)-\xi^i(0)=-\int_0^\tau \Gamma^i_{kl}
\xi^l \frac{d\Pi^k}{ds}ds,
\end{equation}
and, therefore, in the first approximation
\begin{equation} 
|\delta v(y)>=-\Gamma^i_{kl}(0)\xi^l (0)\Delta \Pi^k \frac{\partial \Psi^a}{\partial \Pi^i}|a,y>,
\end{equation}
where
\begin{equation} 
\Delta \Pi^k =\Pi^k(\tau)-\Pi^k(0)=\int_0^\tau \frac{d\Pi^k}{ds}ds.
\end{equation}
This evolution {\it effectivly defines} the map of
the {\it local field dynamical variables} $\xi^i$ in 
CP(N)  to the dynamically shifted states 
$|\Psi +\Delta \Psi>$ in 
original Hilbert space just along a geodesic. 
The mapping of this evolution back to the
full shifted set $\{\Psi^{'m}\}$ like (\ref{PsiPi'}),
by which we can 
identify the spacetime properties of the solution
of a new nonlinear field equation, constitutes a 
continuous renormalization  of the vacuum component 
$\Psi^0$ which
maintains the identity of the physical ground
(vacuum) state. In order to fulfill it one should
use the local coordinates $\{\Pi^i\}$. 
Then one has
\begin{eqnarray} 
|\delta v(y)>=-\Gamma^i_{ml}(0)\xi^l (0)\Delta \Pi^m \frac{\partial \Psi^a}{\partial \Pi^i}|a,y> =
A_m(y)\Delta \Pi^m= \cr \frac{\partial U(y)}{\partial \Pi^m}
\Delta \Pi^m=\frac{\partial u^a}{\partial \Pi^m}|a,y>
\Delta \Pi^m = \frac{\partial S^a_b \Psi^b}{\partial \Pi^m}|a,y>\Delta \Pi^m =\cr
(\frac{\partial S^a_b }{\partial \Pi^m}\Psi^b+
\frac{\partial \Psi^b}{\partial \Pi^m} S^a_b \Psi^b)
|a,y>\Delta \Pi^m.
\end{eqnarray}
Here was introduced the matrix $S^a_b:u^a =S^a_b \Psi^b$ which obeys
the equation
\begin{eqnarray}
\frac{\partial S^a_b }{\partial \Pi^m}\Psi^b+
\frac{\partial \Psi^b}{\partial \Pi^m} S^a_b =
\Gamma^i_{km}(0)\xi^l (0)\frac{\partial \Psi^a}
{\partial \Pi^i},
\end{eqnarray}
in order to write locally linear equation for the 
shifted rate
\begin{eqnarray}
v^a+\delta v^a=-\frac{i}{\hbar}(P_b^a+S_b^a)\Psi^b=
-\frac{i}{\hbar}(P_b^a \Psi^b+u^a).
\end{eqnarray}
The connection between $\Phi^i_\sigma (\Pi,P)$ and $\xi^i$
is simply $\frac{d\Pi^i}{d\tau}=\xi^i=
\Phi^i_\sigma (\Pi,P)\omega^\sigma=
\Phi^i_\sigma (\Pi,P)\frac{d\epsilon^\sigma}{d\tau}$.

5. The ``direct'' comparison of the old and new rates of
the changing of state vectors is possible only in the 
original Hilbert space by {\bf the compensation of the 
geodesic shift with the help of rotations of the 
funtional frame} $\{|a,y>\}$. Therefore one has
\begin{equation}
|v'(y)> =|v(y)+dv(y)>=-\frac{i}{\hbar}[\hat{P}+d\hat{P}]|\Psi(y)>
\label{rate1}
\end{equation}
then
\begin{equation}
-|\delta v(y)>=|dv(y)>=|v(y)+dv(y)>-|v(y)>=-\frac{i}{\hbar}d\hat{P}|\Psi(y)>.
\label{drate}
\end{equation}
That is it is shown in our original Hilbert space $\cal H \rm$ 
the term $|dv(y)>$ arises as an additional rate of a change of the state vector $|\Psi>$
\begin{eqnarray}
|dv(y)>=-\frac{i}{\hbar}d\hat{P}|\Psi(y)>= \frac{\delta U}{\delta \Pi^i}\Delta \Pi^i=
-\Gamma^i_{kl}(0)\xi^l (0)\Delta \Pi^k \frac{\partial \Psi^a}{\partial \Pi^i}|a,y>.
\label{dv} 
\end{eqnarray}
Then $dU=\frac{\delta U}{\delta \Pi^i}d \Pi^i
+\frac{\delta U}{\delta \Pi^{*i}}d \Pi^{*i}$
where
\begin{eqnarray}
A_m(y)=\frac{\delta U}{\delta \Pi^m}=-\hbar \Gamma^i_{km}\xi^k \frac{\partial \Psi^a}{\partial \Pi^i}|a,y>
\label{Am} 
\end{eqnarray}
may be  treated as an ``instantaneous'' 
self-interacting potential of the scalar configuration associated with 
the infinitesimal gauge transformation of the local 
frame  with the coefficients (\ref{connection}).

The spacetime dependence of the distribution of this potential is highly anisotropy relative to directions in
the projective Hilbert space.  
The frequencies of oscillating modes should be extracted
from the formula
\begin{eqnarray}
A_{ij*}(y)=\frac{\delta^2 U}{\delta \Pi^i \delta \Pi^{*j}} 
=-\hbar \{\Gamma^s_{ik}(\frac{\delta \xi^k}{\delta \Pi^{*j}}\frac{\partial \Psi^a}{\partial \Pi^s}+
\xi^k \frac{\delta^2 \Psi^a}{\delta \Pi^i \delta \Pi^{*j}})
+R^s_{kj^*i} \xi^k \frac{\partial \Psi^a}{\partial \Pi^s}\} |a,y>,
\label{R}
\end{eqnarray}
where $R^s_{kj^*i}$ is the Riemannian tensor in CP(N). 
{\bf That is I think the geometric origin of 
the ``elasticity'' of the quantum state which defines 
the spectrum of quantum mass is evoked by the 
counteraction of the curvature of CP(N) and the ``centrifugal'' components of the Christoffel simbol}.
The analysis of this mechanism will be continued.

Strictly speaking, the problem of hierarchy of mass
is not absolutely clear. But we have a some hint 
that {\bf mass (energy) is the threshold of the 
parametric instability of the ``homogeneous'' modes
of oscillations which support the stability of a
system and decay of these modes leads
to the collective reconstruction to self-consistent
boundary. From the phenomenological point of view
this reconstruction is the spacetime motion of 
extended object}. For the relativistic nonlinear scalar field configuration this parametric instability
may be studied in the framework of nonlinear
Klien-Gordon equation
\begin{eqnarray}
\frac{d^2 (\Psi+\Delta \Psi)^*}{d \rho^2} + (3/\rho) \frac{ d(\Psi+\Delta \Psi)^*}{d \rho} + Y(\Delta \Psi,\rho) \cr
+ \alpha^2 (\Psi^*+ \Delta \Psi^* + 
\Psi^* \frac{\partial \Delta \Psi}{\partial \Psi})= 0, 
\label{nlkg}
\end{eqnarray}
\cite {Le7}. 
Then, if our assumption is correct, the behavior 
of the threshold of the parametric instability
on the plane of square of mass--quantum action  $(m^2,\frac{S}{\hbar})$
is similar to the behavior of Regge trajectories.
\vskip 1cm
ACKNOWLEDGEMENTS
\vskip .2cm
I sincerely thank Larry Horwitz for useful discussions 
and critical notes.

\end{document}